\documentclass[superscriptaddress,amsmath,amssymb,aps,pra,twocolumn,floatfix,longbibliography,footinbib]{revtex4-1}

\usepackage{bm,physics}
\usepackage{graphicx}
\usepackage{float}
\usepackage{placeins}
\usepackage{braket}
\usepackage{bbold}
\usepackage{xcolor}
\usepackage{stmaryrd}
\usepackage[colorlinks, linkcolor=blue, citecolor=blue, urlcolor=blue, breaklinks=red]{hyperref}

\newcommand  {\sub} {\mathrm{sub}}

\newcommand {\inphyni} {Universit\'e C\^ote d'Azur, CNRS, Institut de Physique de Nice, France}
\newcommand {\brazil} {Departamento de F\'isica, Universidade Federal de S\~{a}o Carlos, Rod.~Washington Lu\'is, km 235 - SP-310, 13565-905 S\~{a}o Carlos, SP, Brazil}

\begin{document}

\title{Van der Waals dephasing for Dicke subradiance in cold atomic clouds}
\author{Ana Cipris}
\affiliation{\inphyni}

\author{Romain Bachelard}
\affiliation{\brazil}

\author{Robin Kaiser}
\affiliation{\inphyni}

\author{William Guerin}
\affiliation{\inphyni}

\begin{abstract}
We investigate numerically the role of near-field dipole-dipole interactions on the late emission dynamics of large disordered cold atomic samples driven by a weak field. Previous experimental and numerical studies of subradiance in macroscopic samples have focused on low-density samples of pure two-level atoms, without internal structure, which corresponds to a scalar representation of the light. The cooperative nature of the late emission of light is then governed by the resonant optical depth. Here, by considering the vectorial nature of the light, we show the detrimental role of the near-field terms on cooperativity in higher-density samples.
The observed reduction in the subradiant lifetimes is interpreted as a signature of the inhomogeneous broadening due to the near-field contributions, in analogy with the Van der Waals dephasing phenomenon for superradiance.
\end{abstract}

\date{\today}

\maketitle

\section{Introduction}

Collective effects in light-atom interactions are at the focus of intense research, not only for their potential applications to quantum optics and photonics \cite{Asenjo:2017,Chang:2018}, but also for the open fundamental questions in
classical electrodynamics, with the unexpected optical response of a dense atomic gas~\cite{Javanainen:2016, Schilder:2017, Jennewein:2018, Peyrot:2018}, or in mesoscopic physics, such as the Anderson localization of light~\cite{Skipetrov:2016}.
Among the surprising results reported in the last years in the linear-optics regime, it has been shown that the near-field contribution of the dipole-dipole interaction, which becomes important at high density, could prevent Anderson localization of light in 3D~\cite{Skipetrov:2014, Bellando:2014, Skipetrov:2015, Cottier:2019}, and even in 2D \cite{Maximo:2015,Maximo:2019}. In addition, this contribution may be responsible for the failure of the traditional homogenization and mean-field approaches used to describe the steady-state response of a dense atomic gas~\cite{Javanainen:2016, Schilder:2017}. Finally, it also leads to the saturation of the atomic susceptibility at increasing densities~\cite{Andreoli:2020, Corman:2017}.

Regarding the \emph{dynamical} response of a dense atomic cloud, a seminal study by Dicke introduced the concepts of superradiance and subradiance, corresponding to accelerated and slowed-down decay of the excitation for an initially fully inverted system~\cite{Dicke:1954}. In that case too, it has been shown later that the near-field terms of the dipole-dipole interaction break the symmetry properties of the collective states in subwavelength samples, leading to a reduced superradiance~\cite{Friedberg:1972, friedberg:1973, Friedberg:1974}. Practically, the near-field terms induces an inhomogeneous broadening of the eigenvalue spectrum, and the resulting detrimental effect on superradiance is sometimes called `van der Waals dephasing'~\cite{Gross:1982}.

More recently, studies on superradiance and subradiance have focused on the situation of a weakly excited system, either with one quantum of excitation \cite{Scully:2006},  or with a coherent weak driving field (`linear-optics regime') and using very dilute samples~\cite{Guerin:2016a, Araujo:2016, Roof:2016, Weiss:2018}. In these configurations, the subradiant decay rate is governed only by the resonant optical depth of the sample $b_0$, while both the superradiant decay rate and the superradiant excitation dynamics~\cite{Guerin:2019, EspiritoSanto:2020} depend on $b_0$ and on the detuning $\Delta$ of the driving field with respect to the atomic transition. Thus, the macroscopic optical properties of the cloud, rather than the local details, determine its optical response.

In this paper, we investigate the role of the near-field interaction on the subradiant dynamics at increasing densities, yet for clouds larger than the optical wavelength.  More specifically we consider macroscopic disordered samples of intermediate densities, such that both the near-field and far-field terms of the dipole-dipole interaction compete. The main result is that the long subradiant lifetimes are reduced at increasing densities when the coupling of the 
near-field terms are accounted for.
Indeed, near-field terms also induce a strong inhomogeneous broadening on the long-lived part of the eigenvalue spectrum, which is not observed in the scalar light model. The strong analogy between the role of near-field interactions on the superradiant and subradiant dynamics leads us to interpret our observations as Van der Waals dephasing for subradiance.

The paper is organized as follows. In the next section we recall the vectorial- and scalar-light microscopic models for light scattering on ultracold point-like atoms, often called coupled-dipole equations (CDEs). In Sec.~\ref{sec:results}, we show how the subradiant lifetime depends on the cloud properties, and in particular on its density, for each model. The eigenvalue distribution of the corresponding effective Hamiltonian is computed to support the interpretation in terms of inhomogeneous broadening. An Appendix on the impact of close pairs of atoms is included, to make clear the many-atom nature of the subradiance discussed in the body of the paper.

\section{Microscopic model} \label{sec:model}

\subsection{Coupled-dipole equations} \label{sec:vectorial_CDE}

We consider a system of $N$ identical 4-level atoms with ground state $\ket{J_g=0,m_g=0}$ coupled to triple-degenerate excited states $\ket{J_e=1,m_e=0,\pm1}$ by the electric field, with a coupling provided by the dipole transition moment. Atoms are treated as point-like particles with fixed positions $\mathbf{r}_j$, where $j={1,..N}$, and with transition frequency $\omega_{0}=ck_0$, where $k_0$ is the wavevector of the atomic dipole transition. The incident electric field is a plane-wave laser beam described by $E_{\mathrm{in}}=\mathbf{E}_\mathrm{L}\exp(i\mathbf{k}_\mathrm{L}\cdot\mathbf{r}_j-i\omega_\mathrm{L}t)$, characterized by its amplitude $E_\mathrm{L}$, polarization $\hat{\epsilon}_\mathrm{L}$ (with $\mathbf{E}_\mathrm{L}=E_\mathrm{L}\hat{\epsilon}_\mathrm{L}$), wavevector $\mathbf{k}_\mathrm{L}=k_\mathrm{L}\hat{z}$ ($k_\mathrm{L}\approx \omega_{0}/c$ and $\hat{z}$ used as a quantization axis), and frequency $\omega_\mathrm{L}$ detuned by $\Delta=\omega_\mathrm{L}-\omega_{0}$ from the atomic transition.
We here use the spherical basis, with unit vectors $\hat{\mathbf{e}}_\pm=\mp1/\sqrt{2}(\hat{\mathbf{e}}_x \pm \hat{\mathbf{e}}_y), $ and $\hat{\mathbf{e}}_0=\hat{\mathbf{e}}_z$ . Throughout this work we use right-hand circular polarization for the laser beam: $\hat{\epsilon}_L=\hat{\mathbf{e}}_{-1}$.

Our focus is here on a weak driving, when the system presents a linear response to the field, i.e., the linear optics regime. The optical response of the system is given by a set of $3N$ coupled-dipole equations (CDEs)~\cite{Lehmberg:1970a, Lehmberg:1970b, Manassah:2012, Samoylova:2014}:
\begin{align}
    \dot{\beta}_j^{\zeta} &= \left( i\Delta-\dfrac{\Gamma_0}{2} \right){\beta}_j^{\zeta} -i\dfrac{d}{\hbar}\hat{\mathbf{e}}^*_{\zeta}\cdot\mathbf{E}_\mathrm{L}\exp(i\mathbf{k}_\mathrm{L}\cdot\mathbf{r}_j) \nonumber \\ &-\dfrac{\Gamma_0}{2}\sum_{m\neq j} \sum_{\eta}G_{\zeta,\eta}(\mathbf{r}_{jm})\beta_m^{\eta}, \label{vec_CDE}
\end{align}
with $j, m\in \llbracket 1, N\rrbracket$, $\mathbf{r}_{jm}=\mathbf{r}_j-\mathbf{r}_m$ the distance vector between atoms $j$ and $m$. The Green's function reads
\begin{align}
    G_{\zeta,\eta}(\mathbf{r}) & =\dfrac{3}{2}\dfrac{\exp(ik_0r)}{ik_0r} \bigg\{ \left [\delta_{\zeta,\eta}-\hat{r}_\zeta\hat{r}_\eta^* \right]\nonumber \\
  & +\left[\delta_{\zeta,\eta}-3\hat{r}_\zeta\hat{r}_\eta^* \right]\left[\dfrac{i}{k_0r}-\dfrac{1}{(k_0r)^2} \right] \bigg\},
\label{vec_kernel}
\end{align}
with $\zeta,\eta \in (\pm 1,0)$ the spherical-basis components, $d$ the electric-dipole transition matrix element, $\Gamma_0=d^2k_0^3/3\hbar\pi\epsilon_0$ the single-atom decay rate,  and $\hat{r}_\zeta=\hat{\mathbf{e}}_\zeta\cdot \hat{r}$ represents the component of the unit vector $\mathbf{r}/r$ along the direction $\zeta=0,\pm 1$.
Kernel (\ref{vec_kernel}) contains both far-field ($1/r$) and near-field contributions ($1/r^2$ and $1/r^3$). 
The dipole components $\beta_j^\zeta$ represent the amplitude of the induced oscillating atomic dipole, and therefore the scattered electric field at position $\mathbf{r}=r\hat{\mathbf{n}}$ can be obtained from the emission of dipoles:
\begin{equation}
    E_{sc}^{\zeta}(\mathbf{r},t)=-i\dfrac{dk_0^3}{4\pi\epsilon_0}\sum_{j} \sum_{\eta}G_{\zeta,\eta}(\mathbf{r}-\mathbf{r}_{j})\beta_j^{\eta}(t).
\end{equation}

The scalar approximation of the coupled-dipole model, which disregards the vectorial nature of light (i.e., its polarization) and the internal Zeeman structure of the atoms, is obtained by averaging $G_{\zeta,\eta}(\mathbf{r}_{jm})$ in Eq.~\eqref{vec_kernel} over random orientations of the pairs of atoms $j$ and $m$. In the case $\zeta=\eta$, one obtains $\langle \hat{r}_\zeta\hat{r}_\eta^*  \rangle =1/3$, while  $\langle \hat{r}_\zeta\hat{r}_\eta^*  \rangle =0$ for $\zeta\neq\eta$. Therefore, the near-field terms disappear and we obtain the following scalar kernel:
\begin{equation}
    G(\mathbf{r})=\dfrac{\exp(ik_0r)}{ik_0r}.\label{eq:G}
\end{equation}
The atomic dipoles are then described by a scalar $\beta_j$, whose dynamics is given by the scalar CDE:
\begin{equation}
    \dot{\beta}_j=\left( i\Delta-\dfrac{\Gamma_0^\mathrm{(s)}}{2}\right){\beta}_j-\dfrac{dE_L}{\hbar} e^{i\mathbf{k}_\mathrm{L}\cdot\mathbf{r}_j}-\dfrac{\Gamma_0^{\mathrm{(s)}}}{2}\sum_{m\neq j} G(\mathbf{r}_{jm})\beta_m \, ,\label{sca_CDE}
\end{equation}
where the natural decay rate differs by a factor $2/3$ from the vectorial one: $\Gamma_0^{\mathrm{(s)}}=(2/3)\Gamma_0$. Note that the dipole matrix element and the scattering cross section also differ by a factor from the scalar to the vectorial case, showing that the stationary response of the atoms also differ.

In the far-field limit $(r\gg r_j,\ 1/k_0)$, where the electric field is purely transversal, the vectorial kernel is approximated by $G_{\zeta,\eta}(\mathbf{r}-\mathbf{r}_j)\approx\dfrac{\exp(ik_0r)}{ik_0r}(\delta_{\zeta,\eta}-\hat{n}_\zeta\hat{n}_\eta^*)\exp(-ik_0\hat{\mathbf{n}}\cdot\mathbf{r}_j)$, with $\hat{\mathbf{n}}=\hat{\mathbf{r}}/r$. Therefore, the intensity of the scattered light can be computed as $\langle I_{sc} \rangle \propto \langle |E_{sc}|^2\rangle$:
\begin{equation}
    \langle I_{sc}(\mathbf{r},t)\rangle \propto\ \sum_{m,j}e^{-ik_0\hat{\mathbf{n}}\cdot\mathbf{r}_{jm}}\sum_{\zeta, \eta}(\delta_{\zeta, \eta}-\hat{n}_\zeta\hat{n}_\eta^*)\beta_j^{\eta}\beta_m^{\zeta*}.\label{I_vec}
\end{equation}
Note that here $\langle .\rangle$ represents the average over many spatial configurations of atomic positions. In our numerical results, we choose the number of realizations $N_{\mathrm{r}}$ and the number of atoms $N$ such that their product is always the same: $N_r\times N=60000$. Furthermore, considering the azimuthal symmetry (up to the disorder), the obtained intensity is averaged over the azimuthal angle $\phi$.
As for the scalar model \eqref{sca_CDE}, the electric field reduces to
\begin{equation}
    E_{sc}^\mathrm{(s)}(\mathbf{r},t)=-i\dfrac{dk^3}{4\pi\epsilon_0}\sum_{m\neq j} G(\mathbf{r}-\mathbf{r}_{jm})\beta_m(t),
\end{equation}
and the scattered intensity is given by $\langle I_{sc}^\mathrm{(s)}\rangle \propto \langle |E_{sc}^\mathrm{(s)}|^2\rangle$.

To study the decay dynamics, we first compute the steady state of \eqref{vec_CDE} or \eqref{sca_CDE} by solving the linear problem when we set $\dot{\beta}_j^{\zeta}=0$ for any $j$ and $\zeta$ (or $\dot{\beta}_j=0$ for any $j$ in the scalar model). Then this steady-state solution is used as an initial condition to solve the same equation once the laser has been switched off ($E_\mathrm{L}=0$).

\subsection{Atomic sample} \label{sec:at_sample}

We consider a spherical cloud of $N$ motionless atoms with a Gaussian density distribution $\rho_G(\mathbf{r})=\rho\exp(-r^2 /2R^2)$, where $R$ is the rms radius and $\rho=N/(\sqrt{2\pi}R)^3$ is the peak density of the cloud. The resonant optical thickness of such Gaussian cloud is $b_0=3N/(kR)^2$.

Because of the finite atom number that we can simulate (up to several thousands), our study on the interplay between optical thickness and density effects is limited by the range of density for a fixed $b_0$ (and vice versa) that can be achieved. Furthermore, we also impose the condition $R>\lambda$ in order to consider macroscopic samples. As a consequence the dependence of the subradiant lifetime with $b_0$ and $\rho\lambda^3$ can only be studied piecewise. The range of density and on-resonant optical thickness that we study here is $\rho\lambda^3=[0.8 ; 40]$ and $b_0=[2 ; 72]$, respectively.

Note that as one increases the density of the cloud, the probability that close pairs of atoms are generated becomes higher. These pairs result in superradiant and subradiant modes that are characterized by strong energy shifts~\cite{Stephen:1964} (see, e.g., Fig.\,\ref{fig:eigenval_comp}). When driving the system with a significant detuning, some of those pairs may be resonant with the field and consequently be strongly excited. They can then play a significant role in the cloud radiation, despite involving few atoms~\cite{Fofanov:2021}. However, these pairs are expected to be highly sensitive to atomic motion since they involve very short distances; therefore the pairs should not to be relevant for experiments done with thermal clouds. The focus of the present work is on the long-lived \emph{collective} modes, involving many atoms. As a consequence, we hereafter implement a hard-sphere radius for atoms, i.e., an exclusion volume to impose a minimal distance between the atoms, thus minimizing the influence of pairs.

The results presented in Sec.\,\ref{sec:results} have been obtained with a density-dependent exclusion volume defined as $r_{\min} =\rho^{-1/3}/\pi$, since it allows us to explore high densities without introducing significant positional correlations, while efficiently removing the pairs. In Appendix~\ref{sec:exclusion_vol}, we discuss in more details  the possible influence of pairs or of positional correlations on the results presented in the following section.

\section{Subradiant decay dynamics}\label{sec:results}

\subsection{Scaling of subradiance with vectorial light}
\label{sec:scaling_vec}

\begin{figure}[t]
\includegraphics[width=\columnwidth]{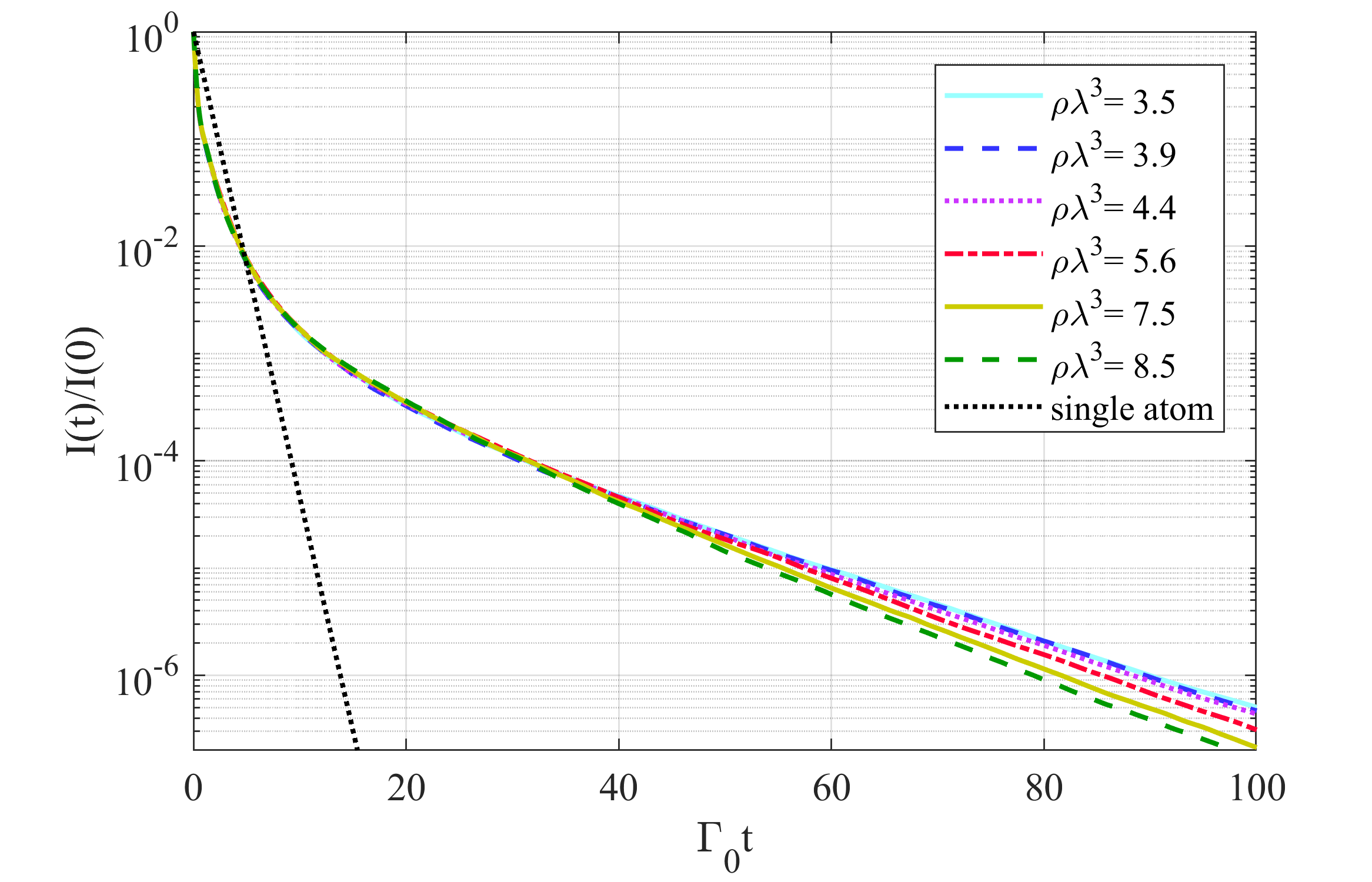}
\caption{Temporal dynamics of the scattered light after the switch-off of the driving field at $t=0$, for a given optical depth $b_0=14$ and for several densities $\rho\lambda^3$. As the density is increased, the decay at late times becomes slightly faster. The total scattered light (all polarizations together) is computed from the vectorial CDE model at $\theta=45^{\circ}$ from the laser propagation axis, after the system has been driven to steady-state with a laser detuned by $\Delta=-15\Gamma_0$ and with circular polarization $\sigma^-$. The exclusion volume is $r_{\min}=\rho^{-1/3}/\pi$.
\label{fig:decay_curve_b0_fixed}}
\end{figure}

\begin{figure}[h]
\centering
\includegraphics[width=\columnwidth]{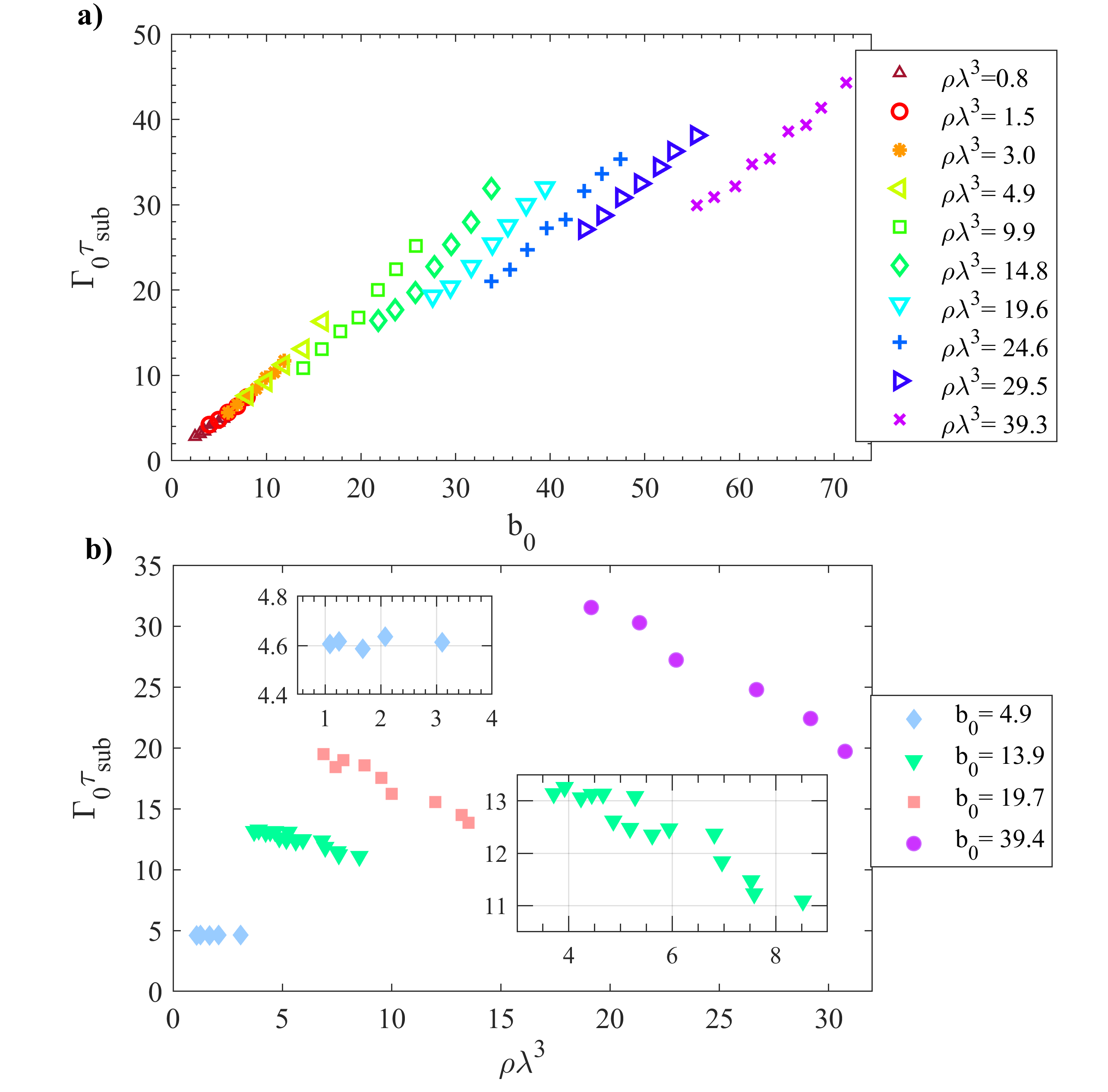}
\caption{Subradiant lifetime $\tau_\sub$ as a function of (a) the on-resonance optical depth $b_0$ for several densities of the sample $\rho\lambda^3$ and (b) the density $\rho\lambda^3$ for several values of $b_0$. 
The insets are close-ups of the two lowest $b_0$ data sets.
The lifetimes $\tau_\sub$ are obtained from an exponential fit of the total scattered light intensity (vectorial model) collected at $\theta=45^{\circ}$ in the fit range $I/I_0=[10^{-6}, 5\times 10^{-6}]$. The parameters of the simulations are the same as in Fig.\,\ref{fig:decay_curve_b0_fixed}.}
\label{fig:tau_b0_rho}
\end{figure}

To understand how near-field terms affect subradiance, we monitor the scattered light intensity, using the vectorial CDEs. The system is first driven to steady-state with a large detuning $\Delta = -15 \Gamma_0$. The subradiant decay rate is then computed from a single-exponential fit of the computed normalized intensity decay $I(t)/I(0)=A\exp(-t/\tau_\sub)$ at late times, according to the procedure used in previous works~\cite{Guerin:2016a, Araujo:2018}. Examples of decay curves are presented in Fig.\,\ref{fig:decay_curve_b0_fixed}.

To illustrate the role of near-field terms on the late-time dynamics, we present in Fig.\,\ref{fig:tau_b0_rho} the subradiant lifetimes for a set of different optical depths and densities. One can see [Fig.\,\ref{fig:tau_b0_rho}(a)] that for lower densities the sets of data points $(b_0,\tau_\sub)$ corresponding to different $\rho\lambda^3$ collapse on the same line. It is also the case for the scalar model in the dilute regime~\cite{Guerin:2016a,Araujo:2018}, and it shows that density effects are negligible for the lowest densities. However, for $\rho\lambda^3\gtrsim5$, the data sets do not collapse any more: higher-density samples present shorter subradiant lifetimes for a given optical depth. This effect of higher densities is even clearer in Fig.\,\ref{fig:tau_b0_rho}(b), where we present the subradiant lifetimes as a function of the density for several values of $b_0$. Again, for the lowest densities there is no visible effect of the density on the long-lived emission, but for $\rho\lambda^3\gtrsim5$ the late lifetimes become shorter with increasing densities.

We have checked that driving the sample with the opposite-sign detuning, as well as with larger detuning, yields the same result. This excludes density-induced collective shifts~\cite{Manassah:2012, Javanainen:2014, Zhu:2016, Jenkins:2016, Jennewein:2018} of the atomic resonance as a source of the observed effect. We have also checked that using other late-time fit intervals (for $I(t)/I(0)<10^{-4}$) leads to the same conclusion (see \cite{Araujo:2018} for the discussion of the fit interval). Finally, we have verified with several off-axis observation angles $\theta=45^{\circ},90^{\circ},135^{\circ},180^{\circ}$ that the conclusion reached from Fig.\,\ref{fig:tau_b0_rho} is independent of the observation angle $\theta$, provided it is outside the forward diffraction lobe, where peculiar effects associated to superradiance may occur~\cite{Kuraptsev:2017}.

\subsection{Comparison with the scalar model}

The density-induced reduction of the subradiant lifetime observed in Figs.\,\ref{fig:decay_curve_b0_fixed}-\ref{fig:tau_b0_rho} occurs in a density regime where the typical distance $r=\rho^{-1/3}$ between atoms is still larger than $1/k_0$. For instance, $\rho\lambda^3=30$ corresponds to $r\simeq 2/k_0$. Therefore both the near-field and far-field terms contribute substantially to the dipole-dipole interaction. It is thus instructive to compare the results with those obtained with the scalar version of the CDEs, where the near-field contribution is absent.

\begin{figure}
\centering
\includegraphics[width=\columnwidth]{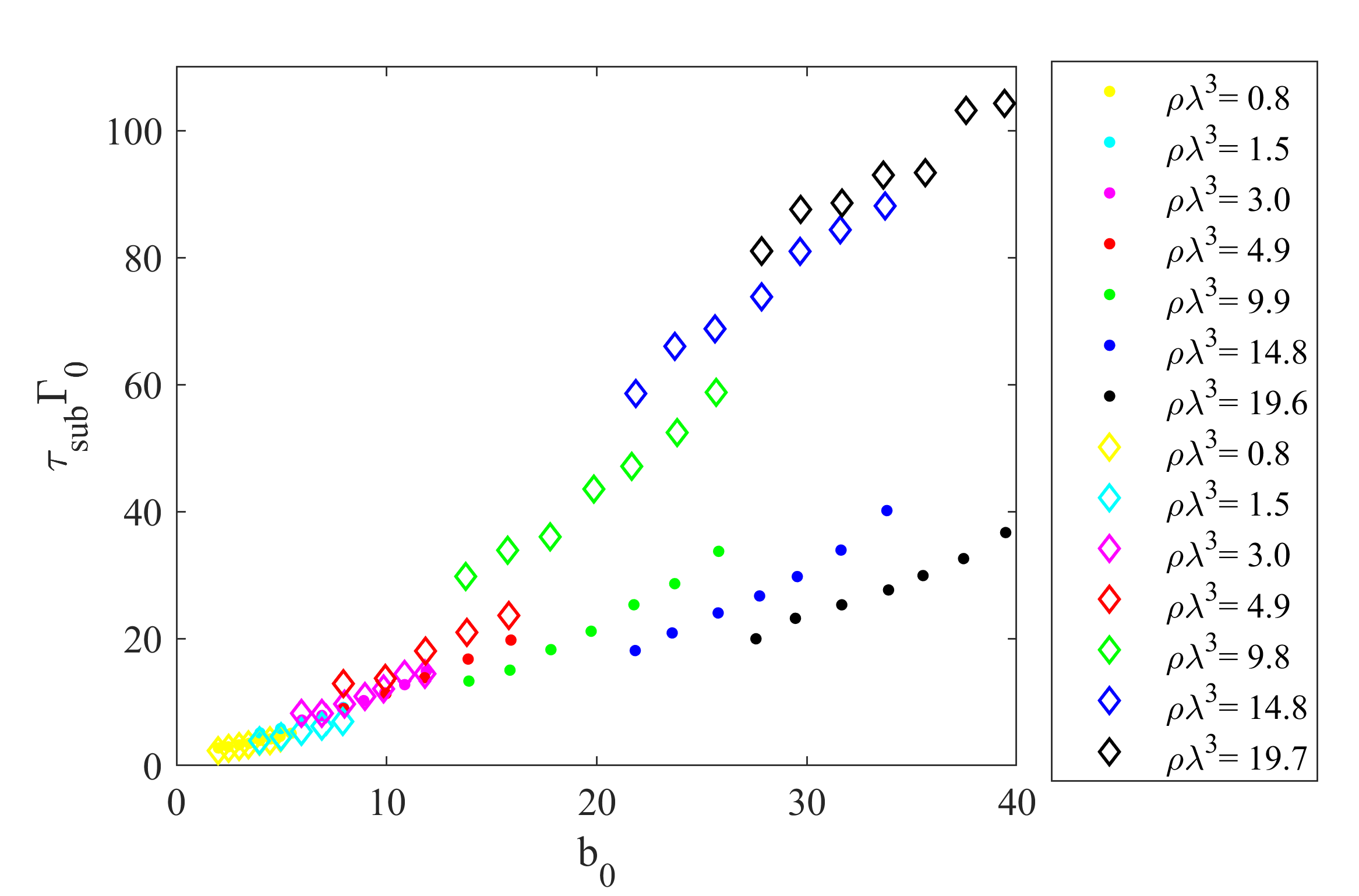}
\caption{Subradiant lifetimes as a function of the resonant optical depth $b_0$ for different densities obtained using the scalar (diamonds) and vectorial models (filled circles). The fit interval is $I(t)/I(0)=[2\times 10^{-7}, 10^{-6}]$. The parameters of the simulation are the same as in Fig.\,\ref{fig:decay_curve_b0_fixed}.}
\label{fig:sc_vs_vec}
\end{figure}

The comparison between the subradiant lifetimes obtained in the two models is presented in Fig.\,\ref{fig:sc_vs_vec}.
Note that in the scalar model, both the resonant optical thickness and the natural decay rate differ by a factor $2/3$ from their vectorial version [$b_0^{(\mathrm{s})}=2N/(k_0R)^2=(2/3)b_0$ and $\Gamma_0^{(\mathrm{s})}=(2/3)\Gamma_0$], so simulations for a given optical thickness and density involve different atom numbers in the scalar and vectorial models. We also restrict ourselves to densities smaller than $\rho\lambda^3 \sim 20$ to avoid the Anderson-localized regime for scalar light~\cite{Skipetrov:2016c}.

As can be observed in Fig.\,\ref{fig:sc_vs_vec}, there is a qualitative difference between the scalar and vectorial subradiant lifetimes at increasing densities. While in the vectorial case we observe a decrease of $\tau_\sub$ with the density, the behaviour is opposite with the scalar model: $\tau_\sub$ increases with $\rho\lambda^3$ for a given $b_0$.
This clearly demonstrates that the reduction of subradiant lifetimes with the density is due to the near-field part of the dipole-dipole interaction.


\subsection{Van der Waals dephasing for subradiant modes}

To gain further insight on these opposite behaviours for scalar and vectorial waves, it is useful to remind the effect of polarizations on the superradiant emission. Originally studied without accounting for near-field terms (i.e., with atoms treated as pure two-level atoms, without internal structure), the superradiant cascade was addressed in two different regimes~\cite{Dicke:1954}. In the case of a subwavelength cloud, it can be assumed that a unique light mode is coupled to the sample, and the cooperativity parameter describing this coupling is the number of particles $N$. Differently, in the case of a macroscopic cloud, such as studied in the present paper, the sample geometry plays a role and the resonant optical thickness $b_0$ is the relevant cooperativity parameter. The importance of near-field terms in dense samples was later recognized, showing that they are detrimental to superradiance~\cite{Friedberg:1972, friedberg:1973, Friedberg:1974, Gross:1982}. Coined ``Van der Waals dephasing'' due to the $1/r^3$ decay they exhibit, the polarization-mixing terms break the symmetry that was central to Dicke's approach, since he would assume the system to decay through a series of symmetric states.

Note that there are two distinct symmetry-breaking effects. The first one rises when the sample size is increased to become comparable or larger than the optical wavelength: even in the scalar light approximation, the atomic dipoles couple to several optical modes, and the cooperativity parameter is then given by the resonant optical thickness $b_0$ rather than the particle number $N$, for macroscopic clouds. This effect was already discussed in the seminal paper by Dicke~\cite{Dicke:1954}. Differently, Van der Waals dephasing corresponds to the strong energy shifts induced by the near-field terms~\cite{Gross:1982}. The inhomogeneous broadening resulting from these terms leads to a reduction of the cooperativity. For subwavelength samples, the reduction of the $N$-factor enhancement characteristic of superradiance stems from the increase of the density rather than a modification of the system size. For these reasons, we here call Van der Waals dephasing the effect on cooperativity of the increase of density, i.e., the rise of the near-field terms, beyond the size effects.

Although we here address the single-excitation regime, the observed behaviour of the subradiant lifetimes is very consistent with the picture developed for the superradiant cascade. For the scalar model, when the density is increased for a fixed $b_0$, the system size reduces (since $R\propto b_0/\rho$); then, the number of particles $N$ is expected to start competing with $b_0$ as the cooperativity parameter. 
This effect can be observed in Fig.~\ref{fig:sc_vs_vec} where, for scalar light (diamond symbols), the subradiant lifetimes increases with the density, for a fixed $b_0$. Note that approaching the localization regime, for which the phase transition is formally reached only for $\rho_c\lambda^3\approx 21$, may also be responsible for this increase in lifetimes~\cite{Skipetrov:2014,Bellando:2014}.
Differently, for vectorial waves (dot symbols), clouds with larger densities present a reduced subradiant lifetimes, confirming that the scenario is very similar to that of the superradiant cascade.

\begin{figure}
\centering
\includegraphics[width=\columnwidth]{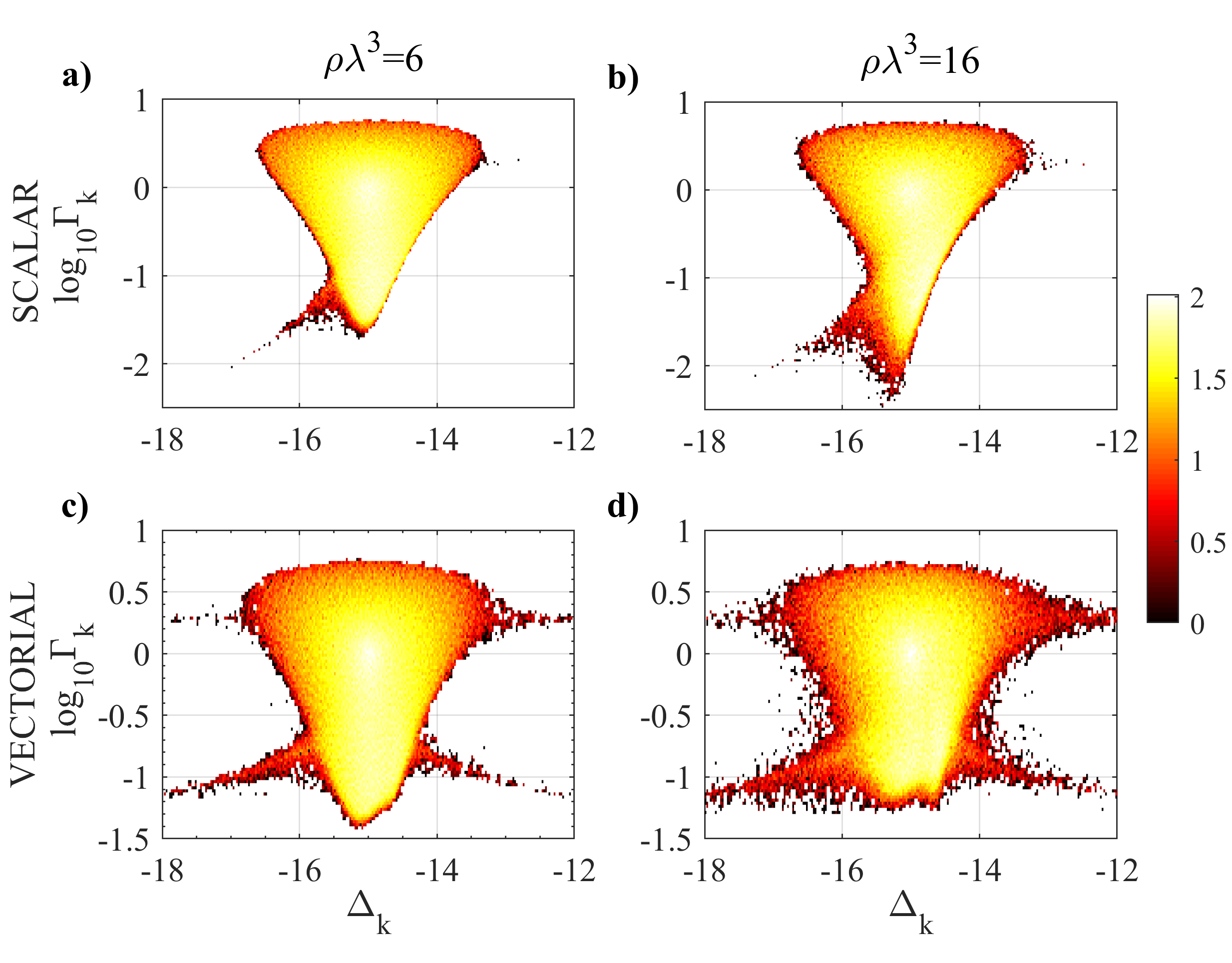}
\caption{Eigenvalue distribution for two different densities, $\rho\lambda^3=6$ (a,c) and $\rho\lambda^3=16$ (b,d) obtained using scalar (a,b) and vectorial (c,d) models. The resonant optical depth is $b_0=20$ for both and the number of realizations is  $N_\mathrm{r}=40$ (a), $N_\mathrm{r}=285$ (b), $N_\mathrm{r}=36$ (c) and $N_\mathrm{r}=255$ (d) such that the number of eigenvalues is the same in all cases. Here there is no exclusion volume and the detuning of the driving laser is $\Delta=-15\Gamma$ (for the sake of consistency with the CDEs, which are written in the laser-rotating frame, the eigenergies are also shifted by $\Delta$). The color code corresponds to the logarithm of the eigenstate density.}
\label{fig:eigenval_comp}
\end{figure}

To extend the analogy with superradiance, we now turn to analyzing the evolution of the spectrum of eigenvalues ($\lambda_k=-\Gamma_k/2+i\Delta_k$). For the single-excitation problem under consideration, it is obtained by diagonalizing the matrices $G_{\zeta,\eta}$ and $G$ given by Eqs.\eqref{vec_kernel} and \eqref{eq:G}. 
The spectrum for scalar light is presented in Fig.~\ref{fig:eigenval_comp} for clouds with respective density $\rho\lambda^3=6$ [panel (a)] and $16$ [panel (b)], both with an optical thickness $b_0=20$. With the increasing density, longer-lived modes appear at the bottom of the distribution, in agreement with the increase of the subradiant lifetimes reported in Fig.\,\ref{fig:sc_vs_vec}. In this case, the broadening of the eigenvalue distribution is very limited.

In contrast, in the presence of near-field terms, the increase in density [from panel (c) to (d)] is characterized by a strong broadening of the spectrum, and the subradiant tail of the distribution is particularly affected. In addition, the longest-lived states disappear with the increasing density for a given $b_0$.
Thus, this broadening is at the origin of the reduction of subradiance. While we here focus on the dynamical features of the scattering, it is interesting to note that inhomogeneous broadening has also been identified as a limiting mechanism for the increase of the refractive index at large densities~\cite{Andreoli:2020}.

\section{Discussion \& Perspectives}

In conclusion, we have reported on the effect of density on subradiance in large atomic clouds. For densities $\rho\lambda^3 \gtrsim 5$, near-field terms induce an inhomogeneous broadening which acts against cooperative effects. This Van der Waals dephasing for subradiance presents very similar features as the one discussed for superradiance \cite{Gross:1982}, where considering smaller samples (in order to increase the density, for a given optical thickness) leads to an increase of cooperativity for scalar waves, and a decrease for vectorial waves.

However, a quantitative difference between subradiance and superradiance is encountered in the densities at which such effects manifest. For the densities studied throughout this paper, superradiance is not substantially affected.
This difference can be attributed by the very different time scales involved in each phenomenon. Indeed, superradiance is very fast, with time scales shorter than $\Gamma_0^{-1}$, so a very strong broadening is required to affect the dynamics over these short times. Differently, subradiance corresponds to modes with lifetimes of many units of $\Gamma_0^{-1}$, making them much more sensitive to the broadening induced by the near-field terms. This analysis is confirmed by the fact that superradiance is more robust than subradiance against inhomogeneous broadening induced by thermal motion~\cite{Weiss:2019, Weiss:2021}.


Apart from the dynamical effects of sub- and superradiance, this inhomogeneous broadening was also shown to prevent Anderson localization of light~\cite{Skipetrov:2014, Bellando:2014} and large refractive indices~\cite{Andreoli:2020}. Inspired by the proposals to recover Anderson localization using a magnetic field to decouple the polarization channels~\cite{Skipetrov:2015, Cottier:2019}, one can imagine enhancing the subradiant lifetimes by applying a similar scheme.



\section*{Acknowledgements}

We thank Igor Sokolov for very fruitful exchanges. Part of this work was performed in the framework of the European Training Network ColOpt, which is funded by the European Union (EU) Horizon 2020 program under the Marie Sklodowska-Curie action, grant agreement No. 721465, and of the project ANDLICA, ERC Advanced grant No.\,832219. We also acknowledge funding from the French National Research Agency (projects PACE-IN ANR19-QUAN-003-01 and QuaCor ANR19-CE47-0014-01), and support from the project CAPES-COFECUB (Ph879-17/CAPES 88887.130197/2017-01). R.\,B. benefited from Grants from S\~ao Paulo Research Foundation (FAPESP, Grants Nos. 2018/01447-2, 2018/15554-5 and 2019/13143-0) and from the National Council for Scientific and Technological Development (CNPq, Grant Nos.\,302981/2017-9 and 409946/2018-4).

\appendix

\section{Exclusion volume} \label{sec:exclusion_vol}

In this appendix we discuss the choice of a density-dependent exclusion volume, introduced to suppress the influence of pairs of very close atoms. First, we check that, with this exclusion volume, pairs indeed do not play a significant role in the decay dynamics. Second, we check that the observed density effect cannot be attributed to the small amount of positional correlations introduced by the exclusion volume.

\subsection{Influence of subradiant pairs}

\begin{figure*}
\includegraphics[width=\textwidth]{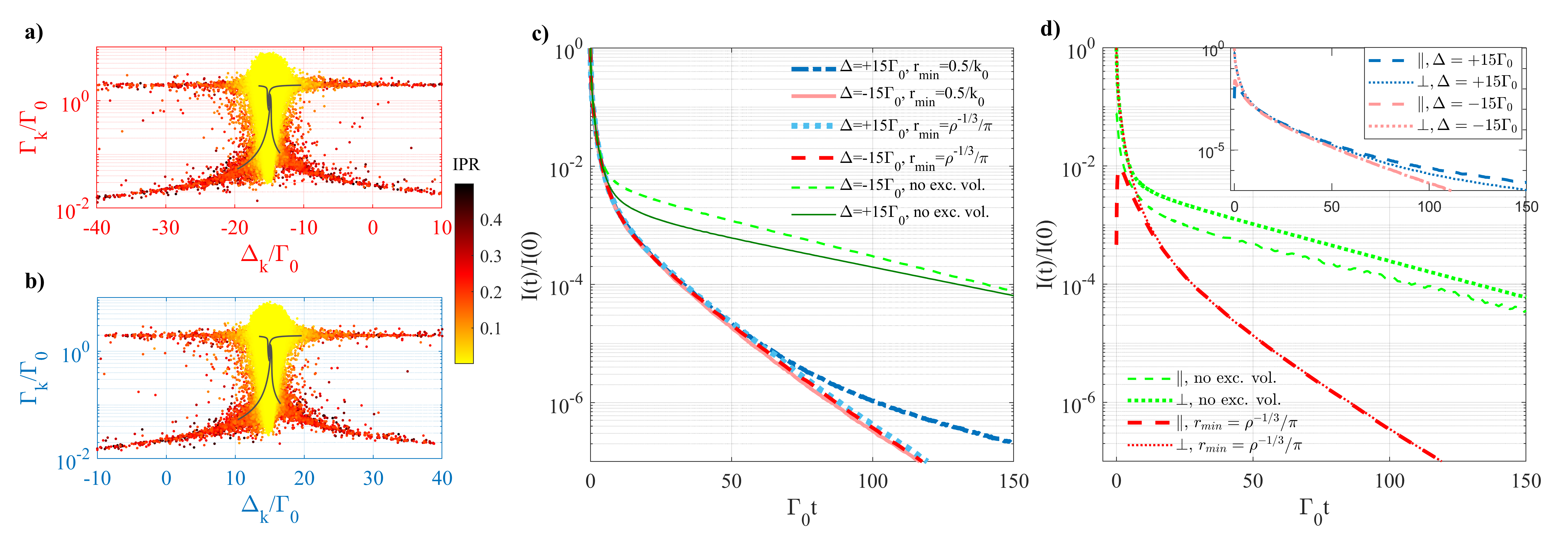}
\caption{Vectorial model. (a,b) Eigenvalue distribution of the collective coupled-dipole modes computed without any exclusion volume and with  (a) a red-detuned driving laser $\Delta=-15\Gamma_0$ and (b) a blue-detuned laser $\Delta=+15\Gamma_0$. The two subradiant pair branches (with IPR $\sim 0.5$) are asymmetric: for a given inter-atomic distance they have different frequencies and lifetimes. The extent of the branches is very long and, depending on the sign of the detuning, one of the branches crosses the resonance $\Delta_k=0$ \cite{Guerin:2017b}. The black lines correspond to the analytical expressions with a cut-off corresponding to the exclusion volume $r_{\min}=\rho^{-1/3}/\pi$. Here $b_0=30, \rho\lambda^3=20$. 
(c) Decay of the scattered light collected at $\theta=45^{\circ}$ for opposite sign detunings and for different exclusion volumes. Without any exclusion volume the entire long-lived dynamics is influenced by pairs, with a noticeable difference between detunings of opposite signs (green lines). On the contrary, for $r_{\min}=\rho^{-1/3}/\pi$, the decay dynamics is independent of the sign of the detuning. With $r_{\min}=0.5/k_0$, only the intermediate dynamics is independent of the sign of the detuning and for very late times the decay for $\Delta = +15 \Gamma_0$ starts to be slower, showing the influence of pairs. (d) Decay of the light scattered at $\theta=90^{\circ}$ decomposed into two polarization channels: parallel ($\parallel$) and orthogonal ($\perp$) to the scattering plane, which is defined by the wave vector of the incoming laser beam and the observation direction. Without any exclusion volume, both superradiant and subradiant parts are partially polarized. With the exclusion volume $r_{\min}=\rho^{-1/3}/\pi$, the superradiant part is polarized while the subradiant part is depolarized. Inset: Two polarization channels of light scattered at $\theta=90^{\circ}$ for the exclusion volume $r_{\min}=0.5/k_0$. For red detuning, the entire subradiant part is depolarized, whereas for blue detuning, light is depolarized at intermediate times and at late times, when there is an influence of pairs, it becomes polarized. For panels (c) and (d): $b_0=16, \rho\lambda^3=10$.}
\label{fig:pairs_vs_nopairs}
\end{figure*}


A safe method to remove the pairs is to use an exclusion radius $r_{\min}=\pi/k_0$, value above which the decay rates of pairs becomes very close to $\Gamma_0$. However this is only appropriate for investigating dilute samples \cite{Guerin:2016a,Araujo:2018} since it is not possible to reach densities higher than $\rho\lambda^3\sim 8$. Here we thus use a less stringent condition with the density-dependent exclusion volume $r_{\min} = \rho^{-1/3}/\pi$ \cite{Moreira:2019}. However it is then necessary to check that the remaining pairs are not responsible for the observed effects.

First of all we illustrate in Fig.\,\ref{fig:pairs_vs_nopairs}(a) and (b) the maximum extent of the pair branches in the eigenvalue distribution for $\rho\lambda^3=10$. The exclusion volume indeed introduces a cut-off in the pair branches [See the black lines in Figs.\,\ref{fig:pairs_vs_nopairs}(a,b)] given by the eigenvalues corresponding to the minimum distance. On the eigenenergy axis $\Delta_k$ the branches stop well before reaching the resonance, showing that the pairs are not particularly well coupled to the driving field in comparison with all the other collective modes \cite{Guerin:2017b}, while on the decay rate axis $\Gamma_k$ it also stops at a value larger than the longest-lived collective modes. We have checked that this observation holds true for all data shown in this paper. Moreover, the subradiant lifetimes obtained from the exponential fit at late times are longer than the lifetimes of the remaining pairs with the cut-off.

Another test which we performed to check the role of pairs is the red-blue asymmetry. For a given distance the pairs results in two subradiant modes with different decay rates. This manifests with a decay dynamics that strongly depends on the sign of the detuning. This effect is shown in Fig.\,\ref{fig:pairs_vs_nopairs}(c), where we plot the decay for positive and negative detuning $\Delta = \pm 15 \Gamma_0$, with and without exclusion volume. For the sake of illustration we also show the results obtained with an exclusion volume defined by a fixed minimal distance $r_{\min} = 0.5/k_0$. Although it removes most of the pairs, at very late time the decay becomes dominated by the pairs for $\Delta = +15 \Gamma_0$, but not for $\Delta=-15 \Gamma_0$. Here a red-blue asymmetry appears. However, for all data presented in this paper (Fig.\,\ref{fig:tau_b0_rho} and \ref{fig:sc_vs_vec}), we have checked that we obtain the same results with the opposite sign of the detuning.

Furthermore, another qualitative difference between collective long-lived modes and subradiant pairs is the polarization of the scattered light. By driving the system with a circular polarization and computing the light scattered at $\theta = 90^\circ$ from the incident direction, we have obtained that subradiant light is fully depolarized when the density-dependent exclusion volume is used. On the contrary, when the scattered light mainly comes from subradiant pairs (as is the case with large detuning, at late times and without exclusion volume \cite{Fofanov:2021}), we obtain a significant imbalance between the two orthogonal polarization channels, see Fig.\,\ref{fig:pairs_vs_nopairs}(d). As previously, we also show in the inset of Fig.\,\ref{fig:pairs_vs_nopairs}(d) the result with $r_{\min} = 0.5/k_0$ for the two signs of the detuning and we observe that at late times, a slight polarization imbalance appears for the positive detuning only, which corroborates our previous observations. Here again, we have checked for all data that the long-lived dynamics is depolarized.

Interestingly, we note that collective long-lived modes yields depolarized light while the superradiant early decay measured at $\theta=90^\circ$ is linearly polarized (orthogonal to the scattering plane). This is fully consistent with an interpretation of superradiance based on a single-scattering event embedded in an effective medium, as discussed recently \cite{Weiss:2021}. Without exclusion volume though, superradiance also has a contribution from pairs, which creates some polarization component in the scattering plane ($\sim 10\%$), as seen in the early decay in Fig.\,\ref{fig:pairs_vs_nopairs}(d).

Finally, we also computed the decay dynamics with a drive on resonance (not shown here), for which we expect to populate the most long-lived collective modes~\cite{Guerin:2017b} rather than the pairs, since the latter are shifted in energy. With or without exclusion volume, we observe a similar reduction on the lifetimes for increasing density.

Together, these tests clearly demonstrate that the reduction of the subradiant lifetime for increasing density cannot be attributed to pair physics.

\subsection{Influence of positional correlations}


\begin{figure}[t]
\includegraphics[width=\columnwidth]{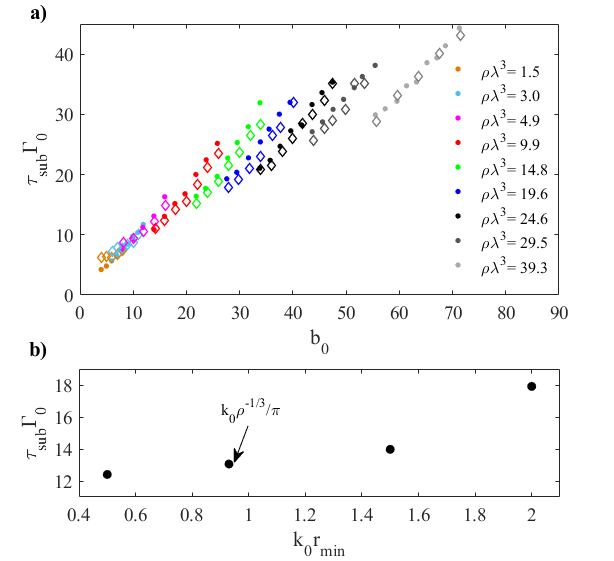}
\caption{Vectorial model. (a) Comparison of subradiant lifetimes obtained with $r_{\min}=\rho^{-1/3}/\pi$ (full circles) and $r_{\min}=0.5/k$ (diamonds) with a red-detuned laser $\Delta=-15\Gamma_0$. (b) Subradiant lifetimes as a function of the exclusion volume $r_{\min}$ for $b_0=16$, $\rho\lambda^3=10$ and a red-detuned laser $\Delta=-15\Gamma$. Lifetimes were obtained in the fit range $I/I_0=[10^{-6}, 5\times 10^{-6}]$.}
\label{fig:tau_compar_EV}
\end{figure}

Imposing an exclusion volume actually induces some correlations in the atomic positions. This correlated disorder, in turn, affects the light scattering properties (see, e.g., \cite{Lax:1951, Rojas:2004, Wang:2020}). In particular, increasing the density at fixed exclusion volume or increasing the exclusion volume at fixed density enhances the positional correlations. In Fig.\,\ref{fig:tau_compar_EV}(a) we show that the lifetimes obtained with two different types of exclusion volume lead to very similar results.

Quantitatively, though, there is a very small influence of the correlations introduced by the exclusion volume, which can be better characterized by plotting the subradiant lifetime as a function of $r_{\min}$ [Fig.\,\ref{fig:tau_compar_EV}(b)]: The subradiant lifetime slightly increases with increasing exclusion volume. Note that this behaviour does not come from pairs since pairs would produce the opposite effect. Moreover, this shows that an increase of the positional correlations yields an increase of the subradiant lifetime. As a consequence, the decrease of the subradiant lifetimes observed for increasing density cannot be attributed to correlations. Therefore, we can conclude that in our work correlated disorder only has a small marginal role on lifetimes and is not causing the reported density effects.

%

\end{document}